# Effects of T-type and L-type calcium currents on synchronized activity patterns in a model subthalamo-pallidal network

Choongseok Park[1*], Leonid L. Rubchinsky[2,3], Sungwoo Ahn[4,5]

[1]Department of Mathematics and Statistics, North Carolina A&T State University, Greensboro, NC, USA 27411
[2] Department of Mathematical Sciences, Indiana University Indianapolis, Indianapolis, IN, USA 46202
[3] Stark Neurosciences Research Institute, Indiana University School of Medicine, Indianapolis, IN, USA 46202
[4] Department of Mathematics, East Carolina University, Greenville NC, USA 27858
[5] Alliance for Brain Stimulation, College of Allied Health Sciences, East Carolina University, Greenville NC, USA 27858

[*] Correspondence:
Choongseok Park
Department of Mathematics and Statistics
North Carolina A&T State University
1601 E Market St
Greensboro, NC 27411, USA
Email: cpark@ncat.edu

## Abstract

Synchronized rhythmic oscillatory activity in the beta frequency band in the basal ganglia (BG) is a hallmark of Parkinson's disease (PD). Recent experiments and theoretical studies have demonstrated the crucial roles of T-type and L-type calcium currents in shaping the activity patterns of subthalamic nucleus (STN) neurons. However, the role of these currents in the generation of synchronized activity patterns in BG networks involving STN is still unknown. In this study, using an STN model incorporating T-type and L-type calcium currents, we examined how these currents shape the patterns of neural activity in the subthalamo-pallidal network, including network dynamics in response to periodic external inputs. The dynamics were studied in relation to the network connectivity parameters - modulated by dopamine (depleted in PD's BG) - and compared with the properties of the temporal patterning of synchronous neural activity previously observed in the experimental studies with Parkinsonian patients. Stronger T-type current enhanced post-inhibitory rebound bursting and expanded synchronized rhythmic activity, reducing the range of intermittent synchrony and increasing resistance to external entrainment. Stronger L-type current prolonged STN bursts, promoted intermittent synchrony over a wide range of input amplitudes, and sustained beta oscillations, suggesting a potential role in the pathophysiology of PD. These results highlight the interplay between intrinsic cellular properties, synaptic parameters, and external inputs in shaping pathological synchronized rhythms in BG networks. Understanding these network mechanisms may advance the understanding of the Parkinsonian rhythmogenesis and further assist in finding ways to modulate and suppress pathological rhythms.

**The basal ganglia in our brain are involved in a series of brain functions including motor behavior. And excessively synchronized rhythmic patterns observed in the basal ganglia are closely related to pathological motor symptoms in Parkinsonian patients. As the only excitatory nucleus in the basal ganglia, the subthalamic nucleus plays an important role in the modulation of activity patterns. Experimental results demonstrated the crucial roles of calcium currents in shaping activity patterns in the STN. In this study, we investigated the effects of two calcium currents on the rhythmic activity patterns within the subthalamo-pallidal network, including activity patterns in response to periodic external inputs. Simulation results provided insight on the interplay between these calcium currents, synaptic parameters related to dopamine depletion typical for Parkinson's disease, and external inputs in shaping pathological rhythms in the network. This study may help advance the understanding of the mechanisms behind motor symptoms of a major neurodegenerative disorder.**

1. Introduction

Parkinson's Disease (PD) is a common neurodegenerative disease, that is characterized by an array of disabling motor and non-motor symptoms. The former includes stiffness in the limbs (rigidity), slowness in the initiation and execution of movement (akinesia and bradykinesia), rest tremor, and postural instability. It is known that PD is closely related to the loss of dopamine in the Basal Ganglia (BG) in the brain. Experimental results have demonstrated a connection between synchronous oscillations of neural activity in the Basal ganglia at the beta band (loosely defined as 10–30 Hz) and the hypokinetic symptoms of Parkinson's Disease (e.g., see reviews Brown, 2007; Hammond et al. 2007; Oswal et al., 2013; Stein and Bar-Gad 2013; Rubin, 2017 as well as recent studies, e.g., Duchet et al., 2021; Yu et al., 2021; Bharti et al., 2025).

This naturally leads to an interest in exploration of potential mechanisms behind the pathological beta-band oscillatory activity. While there may be multiple mechanisms for the beta-band oscillations in the Parkinson's disease, it has long been known (probably at least since Plenz and Kitai, 1999) that excitatory-inhibitory network within BG, consisting of subthalamic nucleus (STN) and external globus Pallidus (GPe), play a key role. This STN-GPe circuit may contribute to the generation of oscillatory synchronized rhythms of Parkinsonian BG (see, e.g., experimental studies in Mallet et al., 2008; Tachibana et al., 2011). Mathematical modeling studies have shown that this excitatory-inhibitory STN-GPe network has an ability to generate a variety of rhythms through rhythmic sequences of recurrent excitation and inhibition independently of or together with inputs to this network (e.g., Terman et al., 2002; Holgado et al., 2010; Park et al., 2011; Dovzhenok and Rubchinsky, 2012; Merrison-Hort and Borisyuk, 2013; Pavlides et al., 2015; Rubin, 2017; Koelman and Lowery, 2019; Ortone et al., 2023; Azizpur Lindi et al., 2024).

Experimental studies of the beta-band synchronized oscillations in different parts of the basal ganglia of Parkinsonian patients during perioperative intervals indicate a highly intermittent, variable temporal structure of these dynamics (Park et al., 2010; Rubchinsky et al., 2012; Ratnadurai-Giridharan et al., 2016; Ahn et al., 2018). This behavior is also consistent with recent observations of the intermittent rhythmicity and incomplete synchrony across cortico-basal ganglia circuits in PD (West et al., 2023; Grennan et al., 2024). The temporal patterning of the synchronized beta oscillations is related to the dopaminergic medication-induced improvements in the motor activity in Parkinsonian patients (Ahn et al. 2018), pointing to its potential relevance to Parkinsonian motor symptoms. Transient synchrony states may affect the motor activity in Parkinson's disease (Tinkhauser et al., 2020). Experimental data seem to suggest that the temporal patterning of synchronized dynamics is characteristic of, and plays an important role in, the dynamics of the basal ganglia networks and associated motor behavior in Parkinson's disease (Cagnan et al., 2019; Rubchinsky et al., 2022; West et al., 2023). It may also play a similar role in the healthy state although the electrophysiological recordings in humans are not possible for ethical reasons.

These experimental studies suggest that it is important to study the different spatiotemporal patterns of activity in the STN-GPe networks and their dynamic origin, which is the subject of this paper. Both cellular and network properties can contribute to these patterns, and, given the nonlinearity of the system, these contributions are likely to be not independent of each other. Thus, it is important to explore how intrinsic properties of neurons within network interact with network parameters to generate these abnormal activity patterns and facilitate transition between normal and pathological states.

Prior experiments demonstrated the crucial roles of T-type calcium (CaT) current, L-type calcium (CaL) current, and hyperpolarization-activated cyclic nucleotide-gated (HCN) current in the activity patterns of STN neurons (Beurrier et al., 1999; Bevan and Wilson, 1999; Bevan et al., 2002a; Hallworth et al., 2003; Wilson et al.,2004; Atherton et al., 2010; Yang et al., 2014). Especially, CaT current and CaL current are essential for slow bursting rhythms observed in STN cells, which, in turn, play a crucial role in the generation of excessively synchronized rhythmic bursting patterns. Based on these experimental results, Park et al. (2021) developed a conductance-based model of STN neuron that encompasses CaT current, CaL current, and HCN current. Using this model, they studied how characteristic activity patterns of STN neuron can be generated through the interplay of these currents using bifurcation analysis (Park et al., 2021).

In the current study, we consider conductance-based STN-GPe network model that incorporates the STN neuron model from (Park et al., 2021) to explore potential mechanisms underlying characteristic synchronous rhythms in STN-GPe network. This study is not specifically aimed at finding the ultimate mechanistic origins of the pathological Parkinsonian beta rhythms, as such an investigation would lie beyond the scope of the methods and approaches of the present study. Rather we are investigating potential mechanisms behind realistic patterns of network activity in a relatively simple network model that nevertheless incorporates kinetics of experimentally relevant currents.

Two network parameters, which may be affected by the lack of the dopamine in the Parkinson's disease, are in the focus of the present investigation. Using these parameters, we first examined whether the considered STN-GPe network could reproduce realistic patterns of the intermittent synchrony and whether the changes in these (presumably dopamine-dependent) parameters result in a transition between lower (presumably healthy) and higher (presumably pathological) synchronous states. We further focused on the roles of CaT and CaL currents in rhythm generation within the STN-GPe network. We also investigated how the network activity patterns can be entrained by the external input signals to STN cells (that may mimic cortical inputs, see, e.g., Tachibana et al., 2011; Ahn et al., 2015, 2016; Pavlides et al., 2015) and examined the effects of CaT and CaL currents on this entrainment process.

The paper is organized as follows: Section 2 presents conductance-based models of STN neuron (subsection 2.1) and GPe neuron (subsection 2.2), an STN-GPe network model with stimulation arrangement (subsection 2.3), and data analysis techniques (subsection 2.4). Section 3 presents the main results. In particular, subsection 3.1 presents effects of CaT and CaL currents on the network dynamics. Subsection 3.2.1 presents the response patterns of the STN-GPe network in response to periodic inputs. In subsection 3.2.2, we investigate the effects of CaT and CaL currents on these response patterns of the STN-GPe network in response to the periodic inputs. In Section 4, we conclude with a discussion.

## 2. Methods

In this section, conductance-based models of STN neuron and GPe neuron are presented. Both models include fast spike-producing potassium and sodium currents ($I_K$ and $I_{Na}$), a calcium dependent voltage-dependent afterhyperpolarization potassium current ($I_{AHP}$), a T-type low-threshold calcium current ($I_{CaT}$), and a leak current ($I_L$). STN neuron also includes a persistent sodium current ($I_{NaP}$), a hyperpolarization-activated cyclic nucleotide-gated current ($I_{HCN}$), an A-type potassium current ($I_A$), and an L-type high-threshold calcium current ($I_{CaL}$). GPe neuron also includes a high-threshold calcium current ($I_{Ca}$). We further describe the network structure and the stimulation arrangement. Finally, we describe data analysis techniques.

### 2.1 STN neuron

STN model follows the one developed in (Park et al., 2021). The differential equation for the membrane potential (*V*) is given by

$$C\frac{dV}{dt} = -I_L - I_K - I_{Na} - I_{AHP} - I_{CaT} - I_{NaP} - I_{HCN} - I_A - I_{CaL} + I_{app0} + I_{app} \tag{1}$$

where $I_L = g_L(V - V_L)$, $I_K = g_K n^4(V - V_K)$, $I_{Na} = g_{Na} m^3 h(V - V_{Na})$, $I_{AHP} = g_{AHP} r^2 (V - V_K)$, $I_{CaT} = g_{CaT} p^2 q (V - V_{Ca})$, $I_{NaP} = g_{NaP}(V - V_{Na})$, $I_{HCN} = g_{HCN} f(V - V_{HCN})$, $I_A = g_A a^2 b (V - V_K)$, $I_{CaL} = g_{CaL} c^2 d_1 d_2 (V - V_{Ca})$. STN cell receives a baseline external input ($I_{app0}$) and additional applied current ($I_{app}$). The units for ionic currents are mA/cm². The dynamics of gating variables, $x \in \{m, h, n, f, a, b, p, q, c, d_1\}$, are described by

$$\frac{dx}{dt} = \frac{x_\infty(V) - x}{\tau_x(V)} \tag{2}$$

For the other two gating variables $x \in \{r, d_2\}$, we have

$$\frac{dx}{dt} = \frac{x_\infty([Ca]) - x}{\tau_x(V)} \tag{3}$$

Voltage dependent steady states ($x_\infty(V)$) for $x \in \{m, h, n, f, a, b, p, q, c, d_1\}$, calcium dependent steady states ($x_\infty([Ca])$) for $x \in \{r, d_2\}$, and voltage dependent time constants ($\tau_x(V)$) for $x \in \{m, h, n, r, a, b, p, q, c, d_1, d_2\}$ are given by

$$x_\infty(V) = \left[1 + \exp\left(\frac{V - \theta_{\infty,x}}{\sigma_{\infty,x}}\right)\right]^{-1}, \quad x \in \{m, h, n, f, a, b, p, q, c, d_1\} \tag{4}$$

$$x_\infty([Ca]) = \left[1 + \exp\left(\frac{[Ca] - \theta_{\infty,x}}{\sigma_{\infty,x}}\right)\right]^{-1}, \quad x \in \{r, d_2\} \tag{5}$$

$$\tau_x(V) = \tau_{0,x} + \tau_{1,x}\left[1 + \exp\left(-\frac{V - \theta_{1,x}}{\sigma_{1,x}}\right)\right]^{-1} + \tau_{2,x} \exp\left(-\frac{V - \theta_{2,x}}{\sigma_{2,x}}\right), \quad x \in \{m, h, n, r, a, b, p, q, c, d_1, d_2\} \tag{6}$$

Voltage dependent time constant $\tau_f(V)$ is given by

$$\tau_f(V) = \tau_{0,f} + \tau_{1,f}\left[\exp(\theta_{1,f} + \sigma_{1,f}V) + \exp(\theta_{2,f} + \sigma_{2,f}V)\right]^{-1} \tag{7}$$

Calcium concentration dynamics are described by

$$\frac{d[Ca]}{dt} = \frac{\epsilon}{2F}(-I_T - I_{CaL}) - K_{Ca}[Ca] \tag{8}$$

Here, [Ca] is the calcium concentration, $\epsilon = 337.1$, F is the Faraday's constant, and $K_{Ca}$ = 0.2/ms is the calcium pump rate.

Maximal conductances (unit: $S/cm^2$) are given by $g_L = 0.9$, $g_K = 57$, $g_{Na} = 49$, $g_{NaP} = 0.003$, $g_{AHP} = 1$, $g_{HCN} = 2$, $g_A = 5$, $g_{CaT} = 20$, $g_{CaL} = 5$. Reversal potentials (unit: $mV$) are given by $V_L = -60$, $V_K = -80$, $V_{Na} = 55$, $V_{HCN} = -43$, $V_{Ca} = 120$. Kinetic parameter values are given in Table 1.

**Table 1.** Values of kinetic parameters in STN neuron model. Units for each parameter values are shown in the first row except $\theta_{\infty,r}$, $\theta_{\infty,d2}$, $\sigma_{\infty,r}$, $\sigma_{\infty,d2}$ whose units are mM.

|   | $\theta_{\infty,x}$ (mV) | $\sigma_{\infty,x}$ (mV) | $\tau_{0,x}$ (msec) | $\tau_{1,x}$ (msec) | $\tau_{2,x}$ (msec) | $\theta_{1,x}$ (mV) | $\sigma_{1,x}$ (mV) | $\theta_{2,x}$ (mV) | $\sigma_{2,x}$ (mV) |
|---|---|---|---|---|---|---|---|---|---|
| m | -40 | -8 | 0.2 | 3 | 0 | -53 | -0.7 | | |
| h | -45.5 | 6.4 | 0.5 | 24.5 | 1 | -50 | -10 | -50 | 20 |
| n | -41.5 | -14 | 0 | 11 | 1 | -40 | -40 | -40 | 50 |
| r | 0.17 (mM) | -0.08 (mM) | 2 | 0 | 0 | | | | |
| f | -75 | 5.5 | 0 | 1 | 0 | -14.59 | -0.086 | -1.87 | 0.08 |
| a | -45 | -14.7 | 1 | 1 | 0 | -40 | -0.5 | | |
| b | -90 | 7.5 | 0 | 200 | 1 | -60 | -30 | -40 | 10 |
| p | -56 | -6.7 | 5 | 0.33 | 200 | -27 | -10 | -102 | 15 |
| q | -85 | 5.8 | 30 | 400 | 100 | -50 | -15 | -50 | 16 |
| c | -30.6 | -5 | 45 | 10 | 15 | -27 | -20 | -50 | 15 |
| $d_1$ | -60 | 7.5 | 400 | 500 | 1 | -40 | -15 | -20 | 20 |
| $d_2$ | 0.2 (mM) | 0.02 (mM) | 3000 | 0 | 0 | | | | |

## 2.2 GPe neuron model

GPe model follows the one originally developed in (Terman et al., 2002) and further modified in (Park et al., 2011). The differential equation for the membrane potential (V) is given by

$$C\frac{dV}{dt} = -I_L - I_K - I_{Na} - I_{AHP} - I_{CaT} - I_{Ca} + I_{gpe} \qquad (9)$$

where $I_L = g_L(V - V_L)$, $I_K = g_K n^4(V - V_K)$, $I_{Na} = g_{Na} m_\infty^3(V)h(V - V_{Na})$, $I_{AHP} = g_{AHP}([Ca]/([Ca] + k_1))(V - V_K)$, $I_{CaT} = g_{CaT} a_\infty^3(V)r(V - V_{Ca})$, $I_{Ca} = g_{Ca} b_\infty^2(V)(V - V_{Ca})$. $I_{gpe}$ is a constant applied current and [Ca] is the concentration of intracellular calcium ions. [Ca] is governed by

$$\frac{d[Ca]}{dt} = \epsilon(-I_{Ca} - I_{CaT} - k_{Ca}[Ca]) \qquad (10)$$

Here, $k_1 = 30$, $k_{Ca} = 3$, and $\epsilon = 0.0055$. The dynamics of gating variables, $x \in \{n, h, r\}$, are described by

$$\frac{dx}{dt} = \emptyset_x \frac{x_\infty(V) - x}{\tau_x(V)} \qquad (11)$$

Voltage dependent steady states ($x_\infty(V)$) for $x \in \{n, h, r\}$ and voltage dependent activation and inactivation variables, ($x_\infty(V)$) for $x \in \{m, a, b\}$, are given by

$$x_\infty(V) = \left[1 + \exp\left(\frac{V - \theta_{\infty,x}}{\sigma_{\infty,x}}\right)\right]^{-1}, \quad x \in \{n, h, r, m, a, b\} \qquad (12)$$

Voltage dependent time constants ($\tau_x(V)$) for $x \in \{n, h, r\}$ are given by

$$\tau_x(V) = \tau_{0,x} + \tau_{1,x}\left[1 + \exp\left(-\frac{V - \theta_{1,x}}{\sigma_{1,x}}\right)\right]^{-1}, \quad x \in \{n, h, r\} \qquad (13)$$

Maximal conductances (unit: $S/cm^2$) are given by $g_L = 0.1$, $g_K = 30$, $g_{Na} = 120$, $g_{AHP} = 30$, $g_{CaT} = 0.5$, $g_{Ca} = 0.1$. Reversal potentials (unit: $mV$) are given by $V_L = -55$, $V_K = -80$, $V_{Na} = 55$, $V_{Ca} = 120$. Kinetic parameter values are given in Table 2.

**Table 2.** Values of kinetic parameters in GPe neuron model.

|   | $\emptyset_x$ | $\theta_{\infty,x}$ (mV) | $\sigma_{\infty,x}$ (mV) | $\tau_{0,x}$ (msec) | $\tau_{1,x}$ (msec) | $\theta_{1,x}$ (mV) | $\sigma_{1,x}$ (mV) |
|---|---|---|---|---|---|---|---|
| n | 0.3 | -50 | 14 | 0.05 | 0.27 | -40 | -12 |
| h | 0.1 | -58 | -12 | 0.05 | 0.27 | -40 | -12 |
| r | 1 | -70 | -2 | 30 | | | |
| m | | -37 | 10 | | | | |
| a | | -57 | 2 | | | | |
| b | | -35 | 2 | | | | |

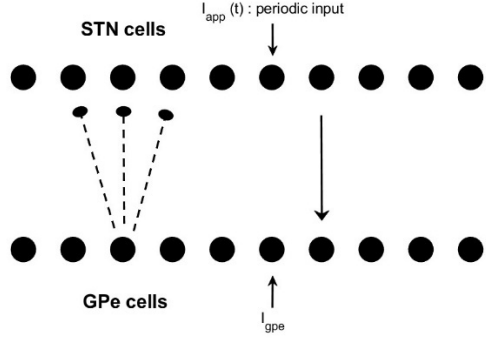

**Figure 1.** STN-GPe network with ten STN neurons and ten GPe neurons. Each STN neuron receives three inhibitory inputs from GPe neurons and each GPe neuron receives one excitatory input from STN neurons. Each GPe neuron also receives an external input ($I_{gpe}$), which reflects a range of dopaminergic deficiency. Each STN neuron may also receive a periodic sinusoidal current input ($I_{app}(t)$), as considered in Section. 3.2.

### 2.3 STN - GPe network

The network architecture is similar to (Park et al., 2011; Ahn et al., 2016; Fig. 1). We use 10 STN neurons and 10 GPe neurons. Each STN neuron receives inhibitory inputs from three GPe neurons and each GPe neuron receives excitatory inputs from one STN neuron (see Equations (16-19)). These synaptic connections are modeled by the first-order kinetic equation describing the fraction of activated channels

$$\frac{d\,s_{x,i}}{dt} = \alpha_x H_{\infty,x}(V_{presyn,x} - \theta_x)(1 - s_{x,i}) - \beta_x s_{x,i}, \quad x \in \{STN, GPe\}, \quad i = 1,2,\ldots,10 \quad (14)$$

where

$$H_{\infty,x}(V) = \left[1 + \exp\left(-\frac{V - \theta_{\infty,x}}{\sigma_{\infty,x}}\right)\right]^{-1}, \quad x \in \{STN, GPe\} \quad (15)$$

The synaptic current from STN to GPe synapse is given by

$$I_{syn,i} = g_{syn,GPe}(V - V_{syn,GPe})s_{STN,i} \quad (16)$$

On the other hand, the synaptic current from GPe to STN synapse is given by

$$I_{syn,1} = g_{syn,STN}(V - V_{syn,STN})(s_{GPe,10} + s_{GPe,1} + s_{GPe,2}), \quad (17)$$

$$I_{syn,i} = g_{syn,STN}(V - V_{syn,STN})(s_{GPe,i-1} + s_{GPe,i} + s_{GPe,i+1}), \quad i = 2,3,\ldots,9 \quad (18)$$

$$I_{syn,10} = g_{syn,STN}(V - V_{syn,STN})(s_{GPe,9} + s_{GPe,10} + s_{GPe,1}), \quad (19)$$

With these synaptic currents, the differential equation for the membrane potential (*V*) of STN neuron is updated to

$$C\frac{dV}{dt} = -I_L - I_K - I_{Na} - I_{AHP} - I_{CaT} - I_{NaP} - I_{HCN} - I_A - I_{CaL} - I_{syn} + I_{app0} + I_{app}(t) \qquad (20)$$

and the differential equation for the membrane potential (*V*) of GPe neuron is updated to

$$C\frac{dV}{dt} = -I_L - I_K - I_{Na} - I_{AHP} - I_{CaT} - I_{Ca} - I_{syn} + I_{gpe} \qquad (21)$$

Synaptic parameter values are given in Table 3.

**Table 3.** Values of synaptic parameters.

|  | $\alpha_x$ | $\theta_x$ | $\beta_x$ | $\theta_{\infty,x}$ | $\sigma_{\infty,x}$ | $V_{syn,x}$ |
|---|---|---|---|---|---|---|
| STN | 5 | -30 | 1 | 39 | 2 | -100 |
| GPe | 2 | -20 | 0.14 | 57 | 2 | 35 |

Two parameters that are varied in this study in order to span a range of dopaminergic deficiency from normal to Parkinsonian states are $I_{gpe}$ (external input to GPe neurons, see Equation (21)) and $g_{syn,STN}$ (synaptic strength from GPe to STN neurons, see Equations (17-19)). For brevity, we will be referring to $g_{syn,STN}$ as just $g_{syn}$. The first one ($I_{gpe}$) affects pallidal excitability. The lower value of this parameter corresponds to stronger pallidal inhibition by external synaptic input, including input coming from striatum. Stronger striato-pallidal inhibition is presumed to be typical in Parkinson's disease due to the lack of dopamine which would otherwise presynaptically suppress striato-pallidal inhibitory transmission The second one ($g_{syn}$) affects the pallido-subthalamic inhibitory synapses. This synaptic transmission is also presumed to be stronger in Parkinson's disease due to similar mechanisms (discussed in Terman et al., 2002; Park et al., 2011; Rubchinsky et al., 2012).

To represent potential input of the beta-band oscillations from the outside of the STN-GPe network (e.g., cortical input to the subthalamic nucleus can potentially transmit this activity), we consider a periodic sinusoidal current input $I_{app}(t)$ (Equation (20)) into all STN cells. Following (Ahn et al., 2016), $I_{app}(t) = A \sin\left(\frac{2\pi\omega}{1000}t\right)$, where *A* is an amplitude and $\omega$ is a frequency in Hz (note that time $t$ is measured in milliseconds in the model). This input potentially represents a cortical input, however, if it has a very large magnitude, it can perhaps be thought of as an input provided by external electrical stimulation (like a low frequency deep brain stimulation in STN).

## 2.4 Data Analysis

### 2.4.1 Transition rates

To analyze the model network activity, we use the data analysis approach similar to that employed in the studies with the analysis of the temporal patterns of synchrony in experimental data (Ahn and Rubchinsky, 2013; Ahn et al., 2014a, 2014b) including those from Parkinsonian patients (Park et al., 2010; Ratnadurai-Giridharan et al., 2016; Ahn et al., 2018; Dos Santos Lima et al., 2020; Targa et al., 2025). This approach follows our previous modeling studies (Park et al. 2011; Dovzhenok et al., 2013; Ahn et al., 2016; Ratnadurai-Giridharan et al., 2017) and is based on the idea of quantifying the transitions between the synchronous and desynchronous states (that is, the vicinity of the synchronized state and the rest of the phase space, see Ahn et al., 2011). We will briefly describe the major steps here, while the references above provide more details. First, we constructed local field potentials (LFPs) as a weighted sum of synaptic inputs to neighboring STN neurons as was done in (Park et al., 2011). Then STN spiking activities (membrane potential time-series) and LFPs were filtered at the beta-band (10-30Hz). Using these filtered data sets, we reconstructed phases of these signals using Hilbert transform (Pikovsky et al., 2001; Hurtado et al., 2004) and created the first-return maps as follows. We first set up a check point for the phase of LFP and then recorded the value of the phase of the spiking signal whenever the phase of the LFP signal crossed this check point from negative to positive values. The resulting sequence of the phases of spiking signal forms the return map. After partitioning the state space of this return map into four equal square regions, the next step is to define synchronized region for this map, where the phase-locked state is placed at the center of one region. Other regions are considered to be desynchronized regions. Lastly, transitions rates between these regions were measured to study the temporal patterns of synchrony. In (Park et al., 2010) experimental data from the STN of Parkinsonian patients and the resulting transition rates were reported and are used here as a reference. If all transition rates in a model are within 0.7 SD of the transition rates found in experimental data, then we regard that the dynamics of a model are similar to that of experiments. In the figures of Results section, these cases were denoted by filled squares.

### 2.4.2 Principal Component Analysis

To quantify the network-wide correlations and identify the regimes of irregular versus synchronized states, we performed Principal Component Analysis (PCA) over the variable *r* (gating variable for $I_{AHP}$ current, see Equation (3)) of model STN cells (Park et al., 2011; Ahn et al., 2016). Although other slow variables would yield similar results, fast variables such as voltage are not appropriate here. This is because even a small difference in the time or shape of a spike will lead to a large number of principle components. This is undesirable, as our goal is to capture synchronization in the slow beta band, which is essentially bursting synchronization, rather than spiking synchronization. The number of principle components that we are evaluating is the number of components in PCA, which capture 80% of the variation of the variable *r* for all STN neurons over 30 sec time windows.

In this study, if the number of principle components is less than 4 (greater than 7, resp.), then it is defined as a synchronized state (an irregular state, resp.). In a synchronized state, activity patterns tend to be a regular spiking or a regular bursting. The lower PCA values are generally associated with higher synchrony and coherence, while higher PCA values are associated with reduced synchrony and coherence. Note that we use the term "irregular" without implying that the spiking necessarily chaotic as this issue is not considered in the present study.

### 2.4.3 Coefficient of Variation

To assess the level of regularity in the activity patterns of STN cells, we used the coefficient of variation (CV) of inter-spike intervals (ISIs) derived from STN cell activity patterns. Here, CV is the standard deviation over the mean of ISIs from STN cells. Generally, CV value increases as the activity transitions from regular spiking to irregular spiking, then to regular bursting with long interburst intervals, and finally to irregular (mixed) bursting rhythms.

There is no strict boundary for classifying activity patterns using CV. In this study, we adopt the following approximate scheme: CV < 0.5 indicates regular spiking; 0.5 < CV < 1 indicates irregular spiking; 1 < CV < 1.5 indicates regular bursting; and CV>2 indicates irregular bursting.

Numerical simulation of the model was performed with XPPAUT (http://www.math.pitt.edu/~bard/xpp/xpp.html). The numerical method was an adaptive-step fourth order Runge-Kutta method with the maximum step size 0.0001 sec, the transient time 5 sec, and total simulation time 35 sec. The data analysis was performed with MATLAB (Mathworks, Natick, MA).

In summary, the data analysis proceeds as follows. First, to compute the transition rates, we filtered the STN voltages and LFPs in the beta band (10–30 Hz) and reconstructed the phases of the signals. We then applied the first-return map analysis to calculate the transition rates (subsection 2.4.1). If the model's transition rates fall within 0.7 SD of the experimental values, we marked those parameter sets using filled squares based on their PCA values (subsection 2.4.2). Second, we used the slow variable $r$ of the model STN cells to compute the number of PCA values that captured 80% of the variance (subsection 2.4.2). These numbers are categorized into four groups (1–3: red; 4–5: green; 6–7: blue; 8–10: black). Third, we computed the CV of ISIs derived from STN cell activity patterns (subsection 2.4.3).

## 3. Results

### 3.1 Effect of CaT and CaL currents on the network dynamics

#### 3.1.1 Baseline dynamics

We consider the baseline dynamics of the model with $g_{CaL}$=5 and $g_{CaT}$=20 (Fig. 2). The dynamics of the network in the space of the dopamine-dependent parameters of $g_{syn}$ (synaptic strength from GPe to STN) and $I_{gpe}$ (external constant input current applied to GPe neuron) are in agreement with prior studies (Park et al., 2011) as the dynamics are getting more coordinated for the larger values of $g_{syn}$ and smaller values of $I_{gpe}$. The numerical simulation results are displayed in Fig. 2A, where the horizontal axis represents $g_{syn}$ and the vertical axis represents $I_{gpe}$. The level of synchrony increases from the upper left corner (presumably corresponding to healthy states) to the lower right corner (presumably representing pathological states associated with lower dopamine levels) as reflected by the number of PCA components: a large number of PCA components (black squares) corresponds to less synchronous dynamics, while a small number of PCA components (red squares) corresponds to more synchronous dynamics.

Similar to the findings of (Park et al., 2011), it is possible to compare the transitions between different regions of the phase space for the model network and the phase space reconstructed from the experimental data (see Methods). The parameter values for which this similarity is reached are marked as filled squares. They tend to correspond to the band of intermittently synchronized activity patterns that diagonally span between the irregular activity region (black squares in the upper left) and the strongly synchronized region (red squares in the lower right). This intermittent band (filled squares), which represents model activity similar to that observed in the microelectrode recordings in the STN of Parkinsonian patients, remains robust with respect to small perturbation to key parameter values. It occupies a relatively broad region along the border of synchronized and irregular dynamics, which may be in agreement with a large variation of symptoms and neural activity in Parkinsonian patients.

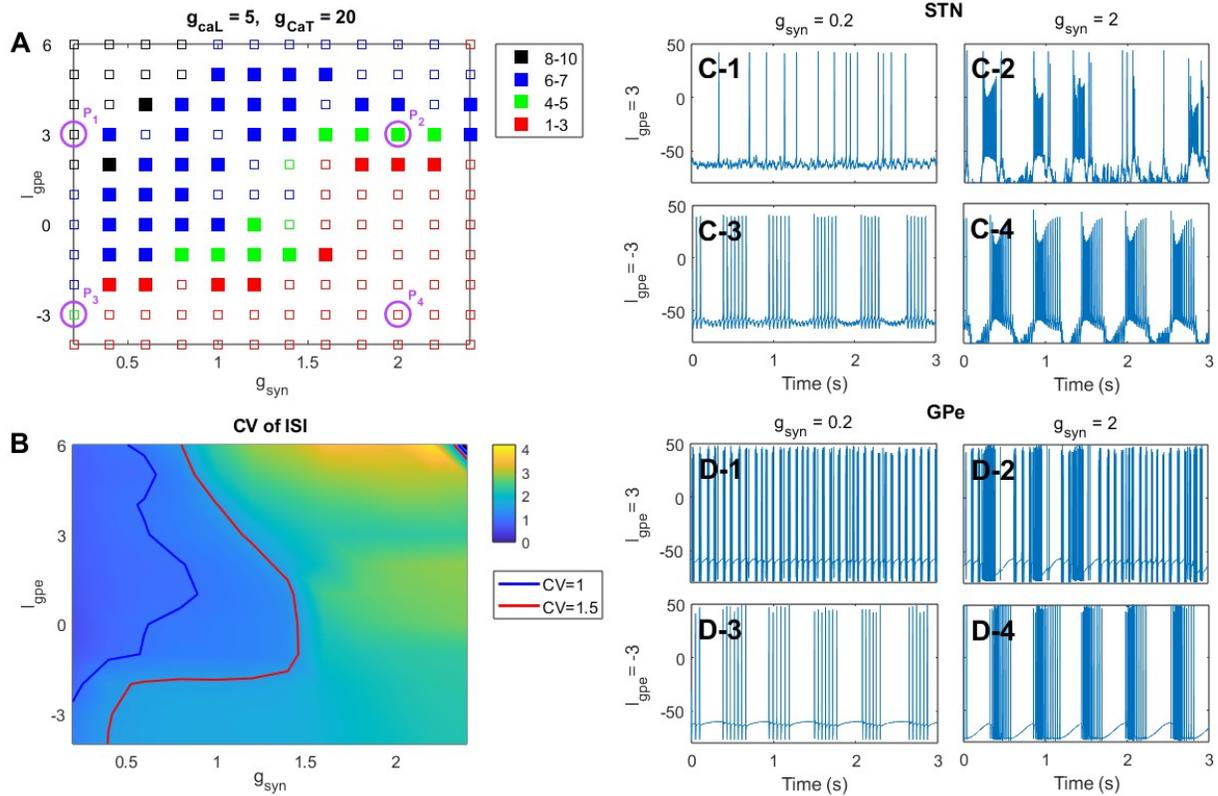

**Figure 2.** Baseline dynamics: $g_{CaL}$ = 5 and $g_{CaT}$ = 20. Two dopamine-dependent parameters, $g_{syn}$ (strength of GPe to STN synapses) and $I_{gpe}$ (external constant current applied to GPe neuron) were chosen for simulation. (A) Plot of numbers of Principle Component Analysis (PCA) components. The color indicates the number of principal components in PCA required to capture 80% of the variability in the calcium dynamics of ten STN neurons. Red represents 1–3 components, green represents 4–5 components, blue represents 6–7 components, and black represents 8–10 components. A large number of PCA components (black squares) corresponds to less synchronous dynamics, while a small number of PCA components (red squares) corresponds to more synchronous dynamics. Filled squares mean that model activity is similar to that observed in the microelectrode recordings in the STN of Parkinsonian patients. (B) Plot of coefficient of variation (CV) of inter-spike intervals of STN cell activity patterns. We used the built-in Matlab *pcolor* (pseudocolor plot) function with interpolation to smooth the color transitions. The blue curve represents CV=1 and the red curve represents CV=1.5. (C-D) Sample activity patterns of STN cells (C) and GPe cells (D) for four sample points (see four purple circles in the panel (A)): (P$_1$) $I_{gpe}$ = 3 and $g_{syn}$ = 0.2, (P$_2$) $I_{gpe}$ = 3 and $g_{syn}$ = 2, (P$_3$) $I_{gpe}$ = -3 and $g_{syn}$ = 0.2, and (P$_4$) $I_{gpe}$ = 3 and $g_{syn}$ = 2.

Fig. 2B illustrates the coefficient of variation (CV) of inter-spike intervals (ISIs) derived from STN cell activity patterns. Generally, CV value increases as the activity transitions from regular spiking to irregular spiking, then to regular bursting with long interburst intervals, and finally to irregular (mixed) bursting rhythms. Based on this, Fig. 2B reveals spiking or regular bursting rhythms on the left side of the parameter space (lower $g_{syn}$), while irregular bursting patterns dominate at the upper right corner.

Four sample points in the parameter space were selected to illustrate several typical activity patterns of the neurons. Those points (marked as purple circles in the panel (A)) are (P$_1$) $I_{gpe}$ = 3 and $g_{syn}$ = 0.2, (P$_2$) $I_{gpe}$ = 3 and $g_{syn}$ = 2, (P$_3$) $I_{gpe}$ = -3 and $g_{syn}$ = 0.2, and (P$_4$) $I_{gpe}$ = 3 and $g_{syn}$ = 2. The corresponding activity patterns are shown for STN cells (Fig. 2C) and for corresponding GPe cells (Fig. 2D).

When external input to GPe neurons $I_{gpe}$ is positive and $g_{syn}$ is weak (P$_1$), neither STN cells nor GPe cells are fully entrained by inputs from their counterparts, leading to irregular rhythms in the STN-GPe network (Fig. 2C-1 and Fig. 2D-1). This irregular activity results in a high number of PCA components (black squares), as shown in Fig. 2A. As $g_{syn}$ increases while $I_{gpe}$ remains positive (P$_2$), the CaT current in STN cells becomes activated by the inhibitory synaptic input from GPe cells. This CaT current enables STN cells to generate bursting rhythms via post-inhibitory rebound (PIR) bursting mechanism, causing the network to generate a mixture of irregular bursting and spiking rhythms (Fig. 2C-2 and Fig. 2D-2). Let us remind here, that filled square means that activity patterns are similar to those in experimental observations in Parkinsonian patients in terms of the organization of the phase space, as noted in (Park et al., 2011).

When $I_{gpe}$ is more negative (P$_3$ and P$_4$), standalone GPe neurons exhibit regular bursting rhythms (data not shown). Under the current network architecture, where STN neurons excite GPe neurons and, in turn, GPe neurons inhibit neighboring STN neurons, STN neurons tend to become entrained by the rhythmic inhibitory input from GPe neurons, while GPe neurons receive rhythmic excitatory input from STN neurons. As a result, the bursting rhythms within the network become more regular as $g_{syn}$ increases (Fig. 2C-3 and Fig. 2D-3).

This entrainment is more prominent when $g_{syn}$ increases as one can see in the lower right region where $I_{gpe}$ is more negative and $g_{syn}$ is large (P$_4$). In that region, STN and GPe cells show fairly regular bursting rhythms and these rhythms are strongly synchronized (Fig. 2C-4 and Fig. 2D-4). Considering that standalone STN neurons display characteristic PIR bursting patterns driven by the CaT current (Park et al., 2021), it can be inferred that the entrainment of STN neurons in STN-GPe network is primarily driven by the CaT current through PIR mechanism. As shown in Park et al 2021, the number of spikes within a burst moderately increases as $g_{syn}$ increases (Fig. 2C-4, P$_4$) since the bursting rhythms are generated by the CaT current via the PIR mechanism.

### 3.1.2 Effects of CaT currents on the network dynamics

We next examine the effect of CaT current on network activity patterns. We consider increased value of CaT current maximal conductance $g_{CaT}$ =30 (from $g_{CaT}$ =20) and observe that synchronous and bursting region (small number of PCA components, red squares) significantly expanded toward upper left corner of the parameter space as CaT current increased (compare Fig. 3A with Fig. 2A). There are two things to note about the effect of increased CaT current on an isolated STN neuron (Park et al., 2021). First, the firing frequency of STN neuron increases monotonically from around 10Hz to 30Hz as $g_{CaT}$ increases. Second, bursting duration of STN neuron during PIR does not change significantly while the number of spikes within a burst increases.

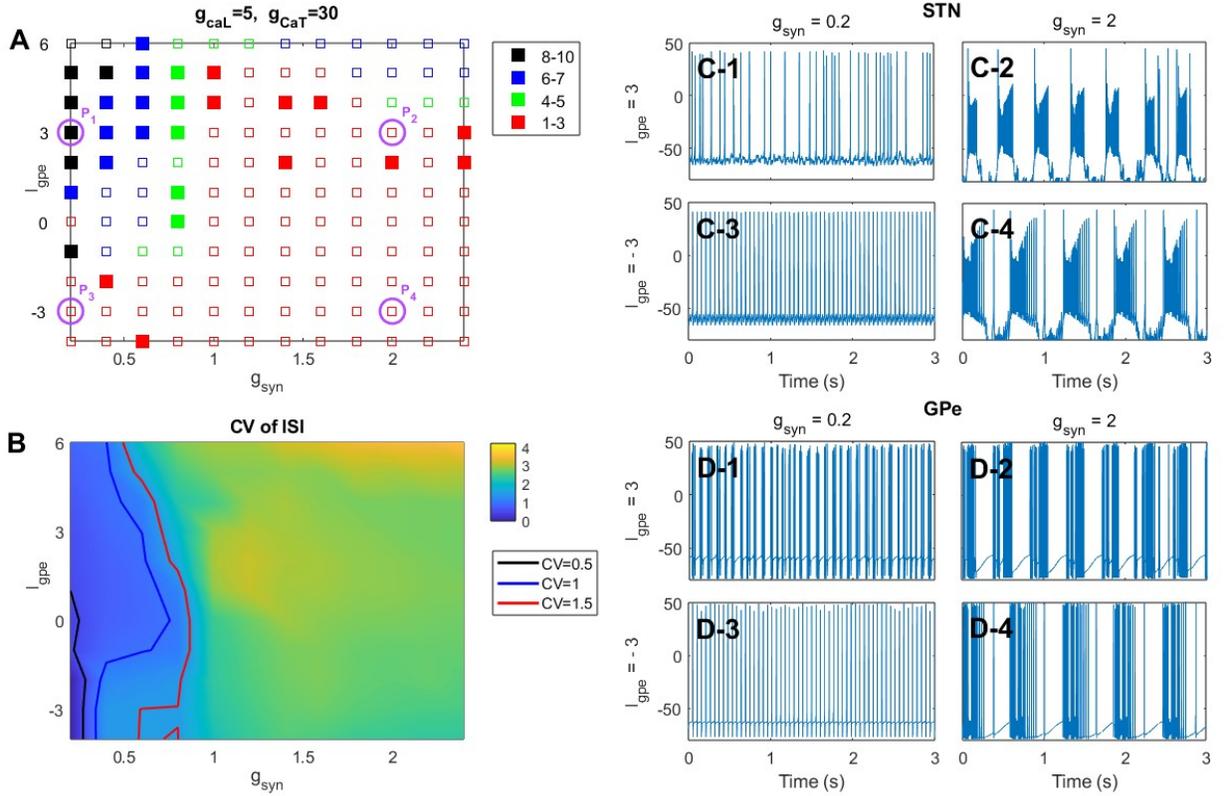

**Figure 3.** Effects of CaT currents on the network dynamics. Maximal conductance of CaT current ($g_{CaT}$) was raised from 20 to 30 while maximal conductance of CaL current ($g_{CaL}$) was kept at 5. Two dopamine-dependent parameters, $g_{syn}$ (strength of GPe to STN synapses) and $I_{gpe}$ (external constant current applied to GPe neuron) were chosen for simulation. (A) PCA plot. The color indicates the number of principal components in PCA required to capture 80% of the variability in the calcium dynamics of ten STN neurons. Filled squares mean that model activity is similar to that observed in the microelectrode recordings in the STN of Parkinsonian patients. (B) CV plot of inter-spike intervals of STN cell activity patterns. The black curve represents CV=0.5, the blue curve CV=1, and the red curve CV=1.5. (C-D) Sample activity patterns of STN cells (C) and GPe cells (D) for four sample points (see four purple circles in the panel (A)): ($P_1$) $I_{gpe}$ = 3 and $g_{syn}$ = 0.2, ($P_2$) $I_{gpe}$ = 3 and $g_{syn}$ = 2, ($P_3$) $I_{gpe}$ = -3 and $g_{syn}$ = 0.2, and ($P_4$) $I_{gpe}$ = 3 and $g_{syn}$ = 2.

Similar to the previous section, we illustrate the activity patterns in four different points of the parameter space to explore the effects of CaT current. The increased spiking frequency in STN neuron is particularly evident in $P_1$ and $P_3$, where $g_{syn}$ is weak (compare Fig. 3C-1 and 3C-3 with Fig. 2C-1 and 2C-3). When $I_{gpe}$ is positive and $g_{syn}$ is weak ($P_1$), the network generates irregular rhythms (Fig. 3C-1 and 3D-1). In contrast, when $I_{gpe}$ is more negative and $g_{syn}$ remains weak ($P_3$), the spiking activity patterns of STN neurons dominate, leading both STN and GPe neurons to exhibit regular spiking (Fig. 3C-3 and 3D-3). Note that when $I_{gpe}$ is more negative, a standalone GPe neuron typically displays bursting rhythms (Park et al., 2011). In other words, GPe neurons in $P_3$ are entrained by the excitatory input from the fast-spiking STN neurons, resulting in higher-frequency spiking rhythms in both neuron types. This phenomenon is clearly reflected in the coefficient of variation (CV) plot (Fig. 3B), where low CV values are observed in the lower left corner, indicating that STN neurons display almost regular spiking activity.

When $g_{syn}$ is sufficiently large, STN neurons respond more reliably to the inhibitory input from GPe neurons and network activity shifts to overall synchronized bursting rhythms. This is visible in the neural activity patterns in $P_2$ and $P_4$. When $I_{gpe}$ is positive and $g_{syn}$ is strong ($P_2$), STN neurons exhibit bursting rhythms (Fig. 3C-2 and 3D-2). Recall that with a lower value of $g_{CaT}$ =20, we previously observed an irregular mixture of spiking and bursting rhythms in STN neurons (Fig. 2C-2). However, with increased $g_{CaT}$, the inhibitory input from GPe neurons becomes strong enough to induce longer PIR bursts in STN neurons, due to the increased "availability" of the CaT current. As a result, the network activity shifts to more coordinated bursting rhythms.

When $I_{gpe}$ is more negative, combined with sufficiently large $g_{syn}$ ($P_4$), STN neurons exhibit longer burst durations driven by the prolonged bursting of GPe neurons (Fig. 3C-4 and 3D-4). Consequently, in the lower right region of the parameter space, network activity patterns become more regular and synchronized (Fig. 3A, 3C-4, 3D-4). This explains the shift of the region with small number of PCA components (red squares). Since PIR burst duration does not change significantly with increased availability of the CaT current (Park et al. 2021), the frequency of the network activity patterns remains relatively unchanged. However, the number of spikes within each burst increases substantially as the availability of CaT current increases. Indeed, in the lower right region, STN neurons display high-frequency, small-amplitude spiking at the onset of PIR bursts (Fig. 3C-4), as observed in (Park et al., 2021).

In summary, within the STN-GPe network, an increase of $g_{CaT}$ promotes more robust and synchronized bursting, especially when combined with strong inhibitory input from GPe neurons. In the bursting regime, smaller number of PCA components reflects this increased regularity. These findings explain why the synchronous, bursting region expands towards the upper left corner of the parameter space, while activity patterns become more regular. Consequently, the band of intermittently synchronized regions shrinks and shifts to the upper left corner. This also indicates that realistic Parkinsonian dynamics for the stronger CaT current corresponds to the upper left corner in the parameter space, which may imply that dopaminergic modulation is decreased by the smaller amount. In other words, CaT affects the degree of the lack of dopaminergic modulation to reach the Parkinsonian state.

### 3.1.3 Effects of CaL currents on the network dynamics

To investigate the effects of CaL current on STN-GPe activity patterns, we increased $g_{CaL}$ from the default value of 5 to 35 (Fig. 4). It is worth to note that in Park et al., 2021, we demonstrated that an isolated STN neuron displayed several CaL-dependent properties: (1) there is an abrupt jump in the frequency of spontaneous tonic firing activity of STN cell as CaL current increases, (2) in hyperpolarization-induced bursting rhythms, CaL current substantially increased burst duration while the inter-burst interval remains relatively unchanged, (3) in PIR, on the other hand, burst duration was substantially increased as $g_{CaL}$ increases, and (4) bursting rhythms are primarily initiated and maintained by the CaT current and once STN cell jumps into bursting regime, CaL current extends the bursting duration as long as intracellular calcium [Ca] is sufficiently available.

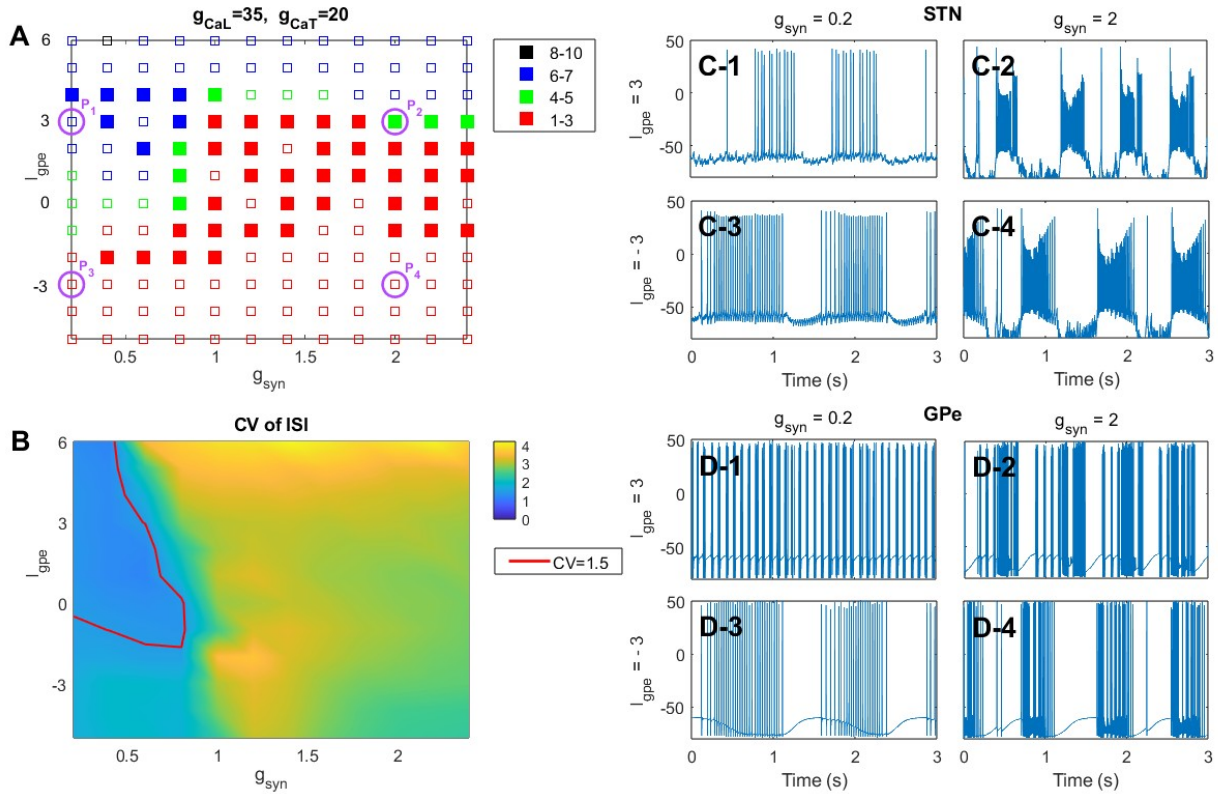

**Figure 4.** Effects of CaL currents on the network dynamics. Maximal conductance of CaL current ($g_{CaL}$) was raised from 5 to 35 while maximal conductance of CaT current ($g_{CaT}$) was kept at 20. Two dopamine-dependent parameters, $g_{syn}$ (strength of GPe to STN synapses) and $I_{gpe}$ (external constant current applied to GPe neuron) were chosen for simulation. (A) PCA plot. The color indicates the number of principal components in PCA required to capture 80% of the variability in the calcium dynamics of ten STN neurons. Filled squares mean that model activity is similar to that observed in the microelectrode recordings in the STN of Parkinsonian patients. (B) CV plot of inter-spike intervals of STN cell activity patterns. The red curve represents CV=1.5. (C-D) Sample activity patterns of STN cells (C) and GPe cells (D) for four sample points (see four purple circles in the panel (A)): ($P_1$) $I_{gpe}$ = 3 and $g_{syn}$ = 0.2, ($P_2$) $I_{gpe}$ = 3 and $g_{syn}$ = 2, ($P_3$) $I_{gpe}$ = -3 and $g_{syn}$ = 0.2, and ($P_4$) $I_{gpe}$ = 3 and $g_{syn}$ = 2.

This distinctive role of CaL current is clearly visible in the current study, particularly when $g_{syn}$ is weak ($P_1$ and $P_3$). Under default (relatively low) value of $g_{CaL}$ =5 (as shown in Fig. 2), we observed either tonic spiking or short bursting rhythms in STN neurons. However, with $g_{CaL}$ = 35, these rhythms were replaced by longer bursting rhythms (Fig. 4C-1 and 4C-3). In $P_3$, even GPe cells exhibited extended bursting rhythms, driven by the excitatory input from STN cells (Fig. 4D-3). When $g_{syn}$ is strong ($P_2$ and $P_4$), on the other hand, network activity patterns are primarily controlled by the bursting rhythms of GPe cells and are reinforced by the response of STN cells. It is known that bursting durations and periods of GPe cell bursting rhythms are limited by its bifurcation structure known as an elliptic burster (Park and Terman, 2010). When $g_{CaL}$ = 35, however, STN cells tend to generate extremely long PIR bursts and this increased excitation from STN cells forces GPe cells to respond even beyond the bifurcation structure of elliptic burster (Fig. 4C-2, 4C-4 and 4D-2, 4D-4).

The region corresponding to small number of PCA components (red squares) expands moderately as $g_{CaL}$ increases and, interestingly, the shape of the band of intermittently synchronized activity patterns (filled squares) was changed from diagonal to horizontal one due to the growth of the region with relatively coordinated dynamics (red filled squares). In other words, from a neurobiological perspective, as far as the experimentally realistic patterns are concerned, there is less dependency on the dopamine-modulated GPe to STN inhibition.

In these red filled squares, network generates longer bursting rhythms, but these rhythms remain intermittently synchronized due to the inhibitory input from GPe cells, which leads to numerous phase slips in spiking. Higher CV values for these rhythms also indicate a greater number of spikes within each burst and longer inter-burst intervals. In other words, CaL current results in almost synchronous longer bursting rhythms but these rhythms render some irregularity. These findings suggest that CaL currents could play a crucial role in beta rhythm generation within the STN-GPe network. Given the fact that burst duration in PIR increases as $g_{CaL}$ increases (Park et al., 2021), it is possible that the prolonged PIR burst duration in STN cells may be a critical factor in pathological rhythm. In other words, the ability of STN neurons to respond to inhibitory inputs over an extended period might be essential for generating pathological beta rhythms.

### 3.1.4. Effect of both calcium currents together on the dynamics of STN-GPe network

We now explore the effects of simultaneous changes in CaL and CaT (considering elevated values of $g_{CaL}$ =35 and $g_{CaT}$ =30). Fig. 5 features the characteristic effects of both parameters at the same time. In $P_1$, for example, the influence of CaL current is more prominent while the influence of the CaT current is more prominent in $P_3$. When $g_{syn}$ is small, through the combining effect of increased CaL and CaT currents, STN cells tend to yield either a longer bursting rhythms or continuous spiking rhythms (Fig. 3 and 4). In $P_1$ where GPe cells show short bursting rhythms, STN cells still yield a longer bursting rhythms since the amount of inhibition from GPe cells is not enough to interrupt STN cell activity patterns. On the other hand, in $P_3$, STN cells yield fast spiking activity patterns driven by the CaT current, which overcome GPe activities although isolated GPe cells show bursting rhythms. That is, GPe cells are entrained by the excitatory input from fast spiking STN cells when $g_{syn}$ is small. This entrainment is clearly visible in CV plot, where the lower left corner is marked by low CV values.

When $g_{syn}$ is sufficiently large, both the number of spikes within a burst (the effect of $g_{CaT}$) and the duration of a burst (the effect of $g_{CaL}$) were significantly increased. STN cells exhibit more regular and synchronized bursting rhythms and the region of small PCA (red squares) expands significantly toward upper left corner (the effect of $g_{CaT}$). On the other hand, there are many red filled squares, which means some added irregularity within synchronized bursting rhythms due to CaL current.

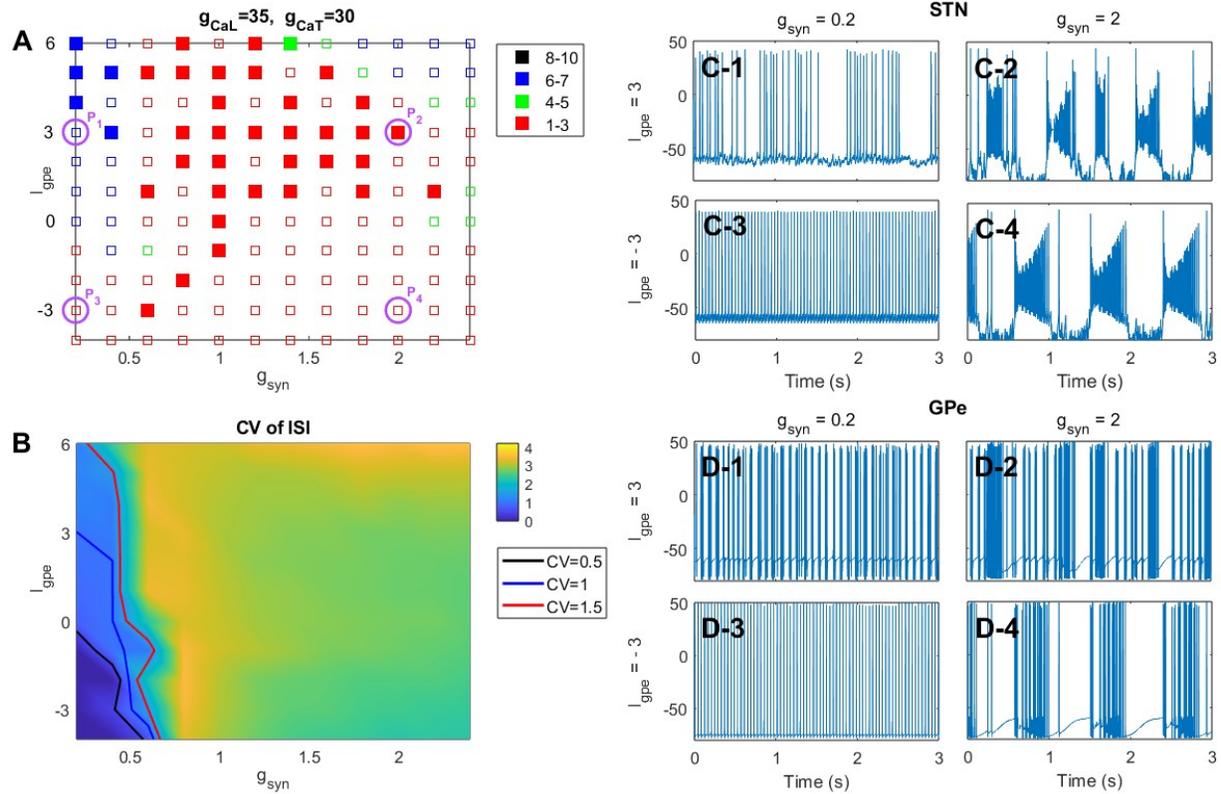

**Figure 5.** Effects of both calcium currents on the network dynamics. Maximal conductance of CaT current ($g_{CaT}$) was raised from 20 to 30 and maximal conductance of CaL current ($g_{CaL}$) was raised from 5 to 35. Two dopamine-dependent parameters, $g_{syn}$ (strength of GPe to STN synapses) and $I_{gpe}$ (external constant current applied to GPe neuron) were chosen for simulation. (A) PCA plot. The color indicates the number of principal components in PCA required to capture 80% of the variability in the calcium dynamics of ten STN neurons. Filled squares mean that model activity is similar to that observed in the microelectrode recordings in the STN of Parkinsonian patients. (B) CV plot of inter-spike intervals of STN cell activity patterns. The black curve represents CV=0.5, the blue curve CV=1, and the red curve CV=1.5. (C-D) Sample activity patterns of STN cells (C) and GPe cells (D) for four sample points (see four purple circles in the panel (A)): ($P_1$) $I_{gpe}$ = 3 and $g_{syn}$ = 0.2, ($P_2$) $I_{gpe}$ = 3 and $g_{syn}$ = 2, ($P_3$) $I_{gpe}$ = -3 and $g_{syn}$ = 0.2, and ($P_4$) $I_{gpe}$ = 3 and $g_{syn}$ = 2.

In Fig. 6, we further investigated the effects of CaT and CaL currents in greater detail by varying the values of the maximal conductances for these currents while selecting four points in the parameter space: A) $g_{syn}$ = 0.8 and $I_{gpe}$ = 3, B) $g_{syn}$ = 2 and $I_{gpe}$ = 3, C) $g_{syn}$ = 0.8 and $I_{gpe}$ = 0, and D) $g_{syn}$ = 2 and $I_{gpe}$ = 0. We varied the values of $g_{CaL}$ and $g_{CaT}$ to examine how STN-GPe activity patterns change. Here, in the baseline case, points A, B, and C are located within the band of intermittently synchronized activity patterns, while point D falls within the lower right synchronized region (Fig. 2A).

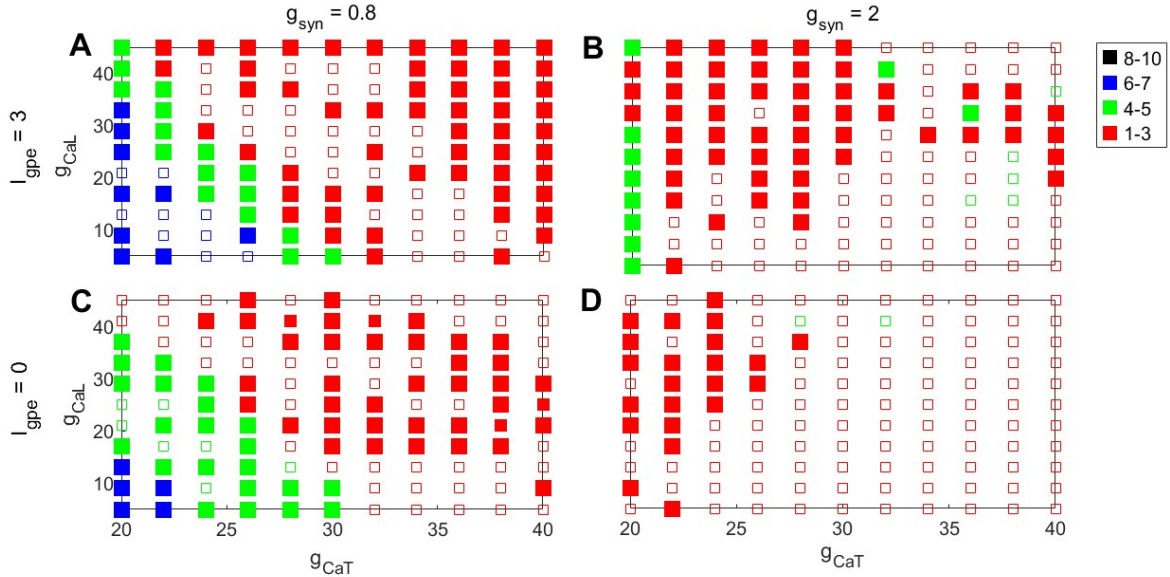

**Figure 6.** Effects of both calcium currents on the network dynamics. Four sample points were chosen to investigate the effects of CaT and CaL currents in greater detail by varying the values of the maximal conductances: (A) $g_{syn}$ = 0.8 and $I_{gpe}$ = 3, (B) $g_{syn}$ = 2 and $I_{gpe}$ = 3, (C) $g_{syn}$ = 0.8 and $I_{gpe}$ = 0, and (D) $g_{syn}$ = 2 and $I_{gpe}$ = 0. In each figure, the horizontal axis shows maximal conductance of CaT current ($g_{CaT}$) and the vertical axis shows maximal conductance of CaL current ($g_{CaL}$). All figures show PCA plot. The color indicates the number of principal components in PCA required to capture 80% of the variability in the calcium dynamics of ten STN neurons. Filled squares mean that model activity is similar to that observed in the microelectrode recordings in the STN of Parkinsonian patients.

As discussed earlier, increasing $g_{CaT}$ generally leads to low-frequency, regular synchronized bursting rhythms (red squares in each panel in Fig. 6). When $g_{syn}$ is large (right column), this transition occurs rapidly (Fig. 6B and 6D). On the other hand, as $g_{CaL}$ increases, activity patterns contain longer bursting and some irregularity is added as a result. Therefore, as $g_{CaL}$ increases, activity patterns may become intermittently synchronized (filled squares). As $g_{CaT}$ increases for fixed $g_{CaL}$ value, activity patterns shift to more regular synchronized bursting rhythms. This is clearly shown when $g_{syn}$ is sufficiently large (transition from red filled squares to red squares in Fig. 6B and 6D). Conversely, when $g_{syn}$ is small and $g_{CaL}$ is large, then we observe a mixture of open red squares and filled red squares (Fig. 6A and 6C).

In summary, increasing $g_{CaT}$ rapidly eliminates intermittently synchronized activity patterns by adding regularity while increasing $g_{CaL}$ moderately expands intermittently synchronized activity patterns by adding irregularity via longer bursting responses. If synaptic strength is weak, then these opposite effects are balanced and result in the transition from more spiking intermittently synchronized regime to more bursting intermittently synchronized regime in a diagonal direction (Fig. 6A and 6C). When synaptic strength is strong, there is an almost instantaneous transition to bursting intermittently synchronized regime (Fig. 6B and 6D). Here strong $g_{CaT}$ eliminates these rhythms. From a standpoint of considering experimentally realistic patterns of activity (filled squares), they are observed for various values of $g_{CaT}$ and $g_{CaL}$, but exhibit different degrees of regularity. Note that a relatively small area of experimentally realistic activity in Fig 6D is because the reference point in the baseline regime is outside of the realistic intermittently synchronized dynamics (see Fig. 2A).

## 3.2 Network dynamics in response to periodic inputs

In this section, we examined the response of the STN-GPe network activity patterns to the external sinusoidal current applied to STN cells in the network (see Methods). We utilized external inputs at three different frequencies (13Hz, 20Hz, and 27Hz, with an idea of considering very low, middle, and high beta-band frequency) with a series of different magnitudes (amplitudes of 3, 6, and 12).

### 3.2.1 Network dynamics with the periodic input currents in the baseline case

Fig. 7 shows the results for the input frequency of 13Hz with an input amplitude of 3 (relatively weak external input to STN cells). Roughly speaking, the responses of STN-GPe network can be categorized into two parts, small $g_{syn}$ case and large $g_{syn}$ case. For small $g_{syn}$, the network exhibits activity patterns similar to the one shown in Fig. 3. Apparently, both the increased $g_{CaT}$ in Fig. 3 and an external input to STN cells result in the increased number of spikes in STN cells and this change in activity patterns of STN cells drives the network activity patterns in both cases. In fact, Fig. 7 shows that STN-GPe network activities at lower left corner of the parameter space tend to be entrained easily by the increased excitatory inputs from STN cells. Continuous and monotonous spiking of STN cell means periodic excitatory input to GPe cells. But, for small $g_{syn}$, GPe cells cannot affect STN cells much. Thus, periodic excitatory input to GPe cells tends to entrain activity patterns of GPe cells. Different from Fig. 3 (which is a no-stimulation case), however, there is a broad region of small CV values at the lower left corner which confirms that this entrainment is occurring due to the periodic input. For large $g_{syn}$, on the other hand, activity patterns are similar to the original patterns shown in Fig. 2. In this case, the effect of external input to STN cells seems not to be significant (potentially due to the small amplitude of input) so that the original activity patterns persist.

Since the number of spikes and/or the spiking frequency of STN cells are determined by the input strength and the entrainment is driven by STN cells, it is natural to expect that the degree of entrainment is determined by the input strength. Numerical simulation shows that the area of entrainment expands from the lower left corner to the upper right corner as the input strength increases (data not shown). If STN cells render periodic spiking patterns, then the corresponding CV values will be very low. Now, if strong periodic excitatory inputs from STN cells entrain GPe cells, then the expansion of entrained region can be seen in CV plot indirectly. In fact, numerical simulation also shows that the region for low CV values expands from the lower left corner to the upper right corner (data not shown). Accompanying this expansion of the entrained region, the region representing intermittently synchronized activity shrinks and shifts toward upper right corner of the parameter space as the amplitude of periodic input increases. In this upper region, irregular bursting rhythms driven by strong inhibitory input from GPe cells persist until external input strength is sufficiently strong.

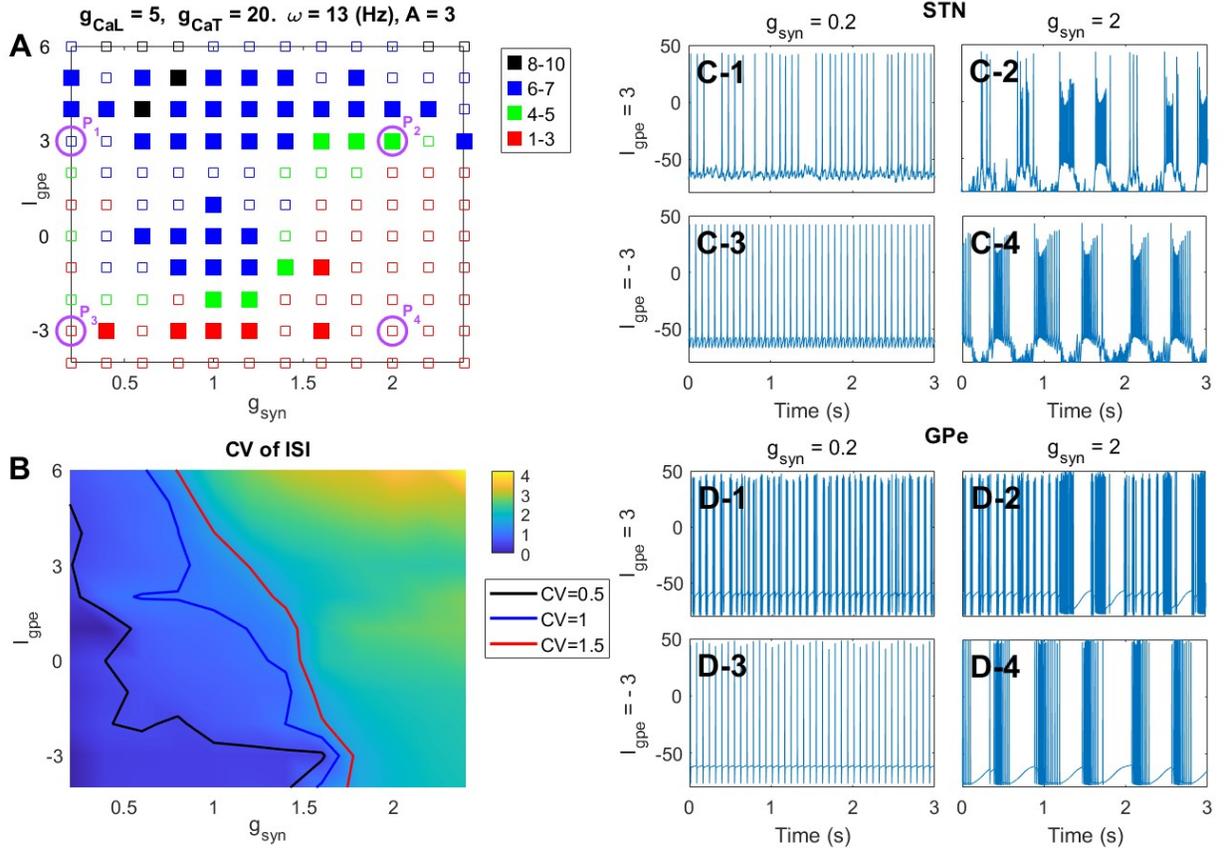

**Figure 7.** Network dynamics with the periodic sinusoidal input currents (see Methods) in the baseline case ($g_{CaL}$ =5, $g_{CaT}$ =20). This figure shows the results for the low input frequency of 13Hz with an input amplitude of 3. Two dopamine-dependent parameters, $g_{syn}$ (strength of GPe to STN synapses) and $I_{gpe}$ (external constant current applied to GPe neuron) were chosen for simulation. (A) PCA plot. The color indicates the number of principal components in PCA required to capture 80% of the variability in the calcium dynamics of ten STN neurons. Filled squares mean that model activity is similar to that observed in the microelectrode recordings in the STN of Parkinsonian patients. (B) CV plot of inter-spike intervals of STN cell activity patterns. The black curve represents CV=0.5, the blue curve CV=1, and the red curve CV=1.5. (C-D) Sample activity patterns of STN cells (C) and GPe cells (D) for four sample points (see four purple circles in the panel (A)): (P$_1$) $I_{gpe}$ = 3 and $g_{syn}$ = 0.2, (P$_2$) $I_{gpe}$ = 3 and $g_{syn}$ = 2, (P$_3$) $I_{gpe}$ = -3 and $g_{syn}$ = 0.2, and (P$_4$) $I_{gpe}$ = 3 and $g_{syn}$ = 2.

For other frequencies (20 Hz and 27 Hz), the overall trend of entrainment is qualitatively similar to the results for an input frequency of 13 Hz (Fig. 8). However, the initiation of entrainment is frequency-dependent. As an example, entrainment in the lower left corner for an input frequency of 13Hz begins with relatively small input strength (Fig. 8 upper row). On the other hand, for an input frequency of 27 Hz, entrainment occurs only at relatively large input strength. In general, for higher input frequency, it requires stronger input amplitude for the entrainment. For sufficiently large input strength, nearly all regions were entrained by external inputs eventually.

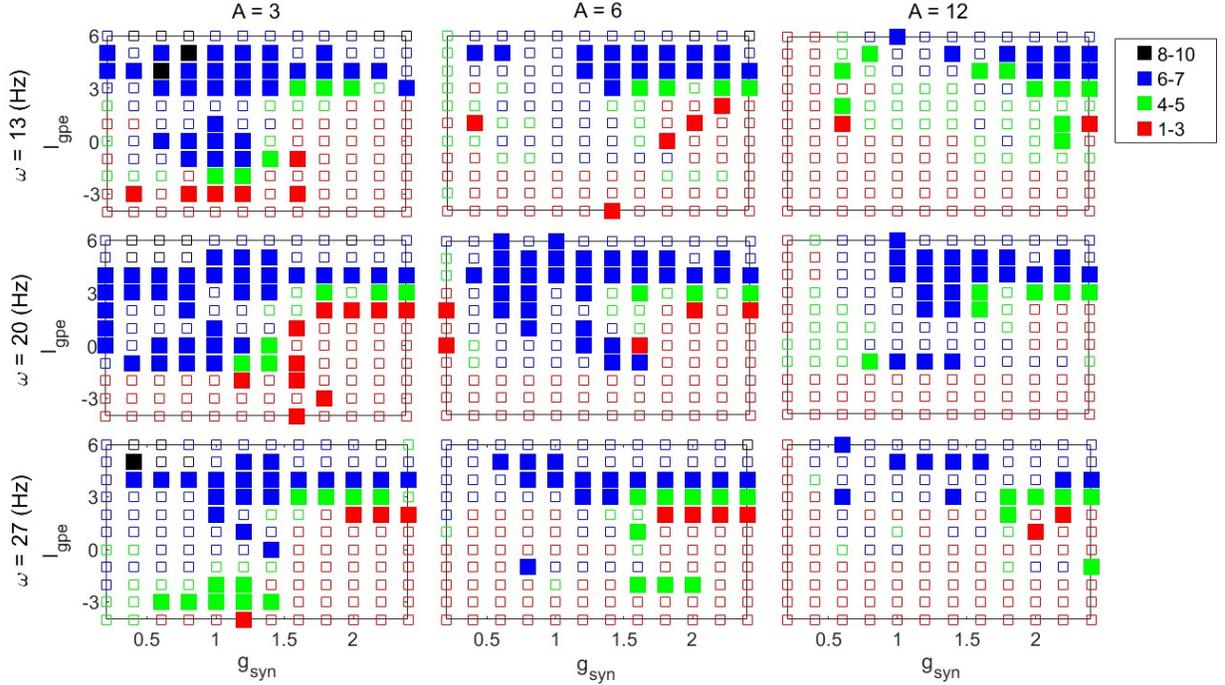

**Figure 8.** Network dynamics with the periodic input currents in the baseline case. This figure shows PCA plots for external inputs at three different frequencies (ω = 13Hz, 20Hz, and 27Hz, with an idea of considering very low, middle, and high beta-band frequency) with a series of different amplitudes (*A* = 3, 6, and 12). All figures show PCA plot. The color indicates the number of principal components in PCA required to capture 80% of the variability in the calcium dynamics of ten STN neurons. Filled squares mean that model activity is similar to that observed in the microelectrode recordings in the STN of Parkinsonian patients.

### 3.2.2 Effect of both calcium currents on the network dynamics in response to the periodic inputs

In this section, we examined the effects of CaT current ($g_{CaT}$ = 30) and CaL current ($g_{CaL}$ = 35) on the entrainment of network activity patterns driven by external input with the frequency of 20Hz, that is, in the middle of the beta frequency band.

Consider stronger CaT current ($g_{CaT}$ = 30). The no input case is in Fig. 3. As the input strength increases (see Fig. 9 upper row), the lower left region in the parameter space has networks that respond to external input and become entrained even with a relatively small input amplitude of 3. Here we note that the expansion rate of the entrained region with respect to input amplitude is faster than the default value of $g_{CaT}$ =20 (Fig. 8, middle row). The expansion rate of the entrained region with respect to input amplitude also depends on the frequency as in Fig. 8 (data not shown). In the region where the network activity patterns are close to the experimentally observed Parkinsonian ones (filled squares in Fig. 9), we observe that this region is getting smaller since the network is moving into highly regular and synchronized state as the amplitude increases (too synchronous even for Parkinsonian neurophysiology).

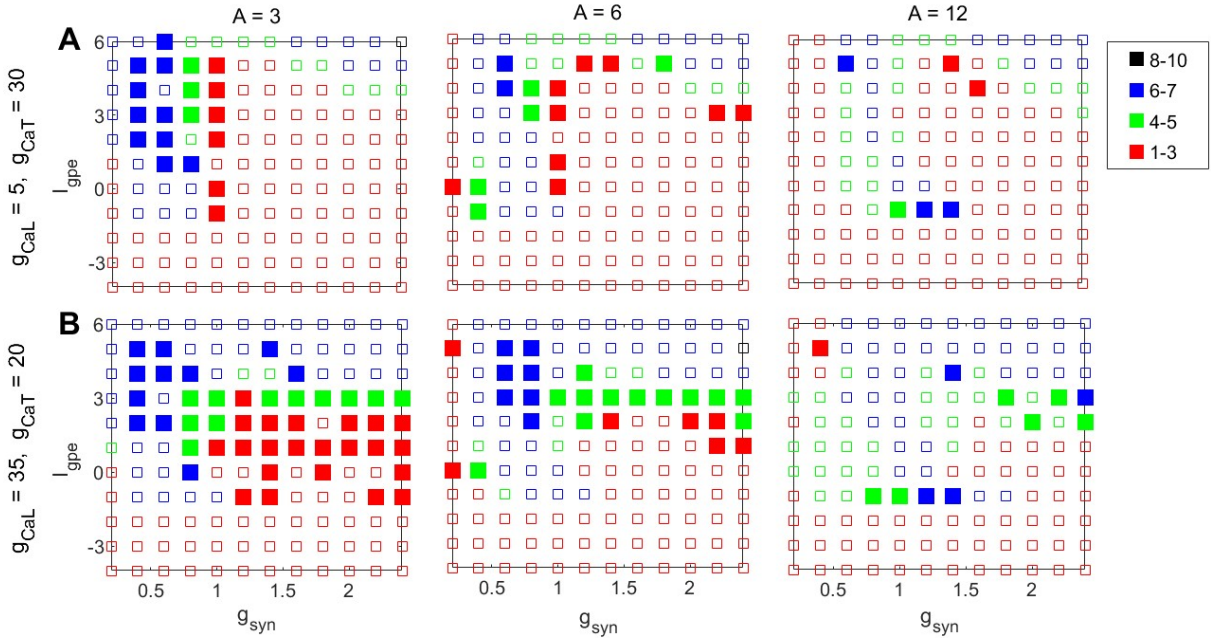

**Figure 9.** Effect of both calcium currents on the network dynamics in response to the periodic inputs. In all figures, external input frequency is 20 Hz with amplitudes ($A = 3, 6,$ and $12$). (A) Maximal conductance of CaT current ($g_{CaT}$) was raised from 20 to 30 while $g_{CaL} = 5$ (default value). (B) Maximal conductance of CaL current ($g_{CaL}$) was raised from 5 to 35 while $g_{CaT} = 20$ (default value). All figures show PCA plot. The color indicates the number of principal components in PCA required to capture 80% of the variability in the calcium dynamics of ten STN neurons. Filled squares mean that model activity is similar to that observed in the microelectrode recordings in the STN of Parkinsonian patients.

Now, let us consider stronger CaL current ($g_{CaL} = 35$) (Fig. 9 lower row). There is a broad region of intermittently synchronized activity patterns persisting across a relatively wide range of input amplitudes. As stated earlier, CaL current results in longer bursting rhythms with some irregularities and this slows down the rate of entrainment as the input strength increases. However, even though the larger CaL case (in contrast with larger CaT case above) preserves a large domain in the parameter space with irregular, less synchronized dynamics, these dynamics are becoming less and less experimentally realistic with rising amplitude of the input.

In summary, CaT current promotes the entrainment driven by external inputs, but CaL current does not affect much. This is because CaT current promotes periodic continuous spiking patterns with increased spiking frequency, but CaL current promotes longer bursting rhythms with increased irregularity.

**Discussion**

As described in the Introduction, multiple experimental studies have demonstrated the importance of the patterns of beta-band rhythmicity in the subthalamo-pallidal networks. The results therein have led to the development of multiple models for this network to study the dynamics of these patterns, including the models that utilized the matching of the phase space (Park et al., 2011; Ahn et al., 2016). As the only excitatory nucleus in the BG that receives massive cortical projections, STN is believed to play a crucial role in shaping BG dynamics (Bevan et al., 2002b; Kühn et al., 2009; Hirschmann et al., 2011; Rubin, 2017; Huang et al., 2021; Tai, 2022; Pasquereau and Turner, 2023). Thus, it is important to understand how the membrane properties of STN are affecting these rhythmic dynamics.

Experimental studies have demonstrated the crucial roles of CaT current and CaL current in the activity patterns of STN neurons (Beurrier et al., 1999; Bevan and Wilson, 1999; Bevan et al., 2002a; Hallworth et al., 2003; Wilson et al.,2004; Atherton et al., 2010; Yang et al., 2014). The impact of these two calcium currents on the dynamics of the isolated STN neuron was studied in (Park et al., 2021), which showed how these currents are important for characteristic beta-band bursting activity patterns. It was shown that CaT current enables STN neurons to generate activity patterns under hyperpolarizing stimuli and CaL current enriches and reinforces these dynamics. These two currents interact synergistically enabling STN neurons to respond to hyperpolarizing stimuli in a salient way.

In this study, we explored the activity patterns of the STN-GPe network using the STN model that involves these CaT and CaL currents (as developed in Park et al., 2021), focusing on the critical roles of these currents in rhythm generation within the network. Furthermore, we investigated the entrainment of the network under external periodic inputs given to STN cells and analyzed the effects of CaT and CaL currents on this entrainment.

We considered two parameters relevant to dopamine depletion characteristic for Parkinson's disease: $I_{gpe}$ (external input to GPe cell) and $g_{syn}$ (synaptic strength of GPe to STN connections). Similar to the earlier findings (Park et al., 2011), this parameter space was roughly divided into two subregions: a region for irregular activity patterns ("healthy states") and a region for synchronized bursting rhythms ("pathological states"). Additionally, there was a band of intermittently synchronized activity patterns resembling those observed in Parkinson's disease in terms of the fine temporal patterns of synchrony. Overall, synchronized bursting dynamics for higher values of $g_{syn}$ may fit well with the observation (John et al., 2023) of the robust bursting in square-wave bursters enhanced by slow negative feedback (STN excites GPe and GPe inhibits STN, so strong $g_{syn}$ may effectively create negative feedback for STN).

When CaT current was increased, STN cell tended to respond more easily to inhibitory input from GPe cells, generating bursting rhythms via Post-Inhibitory Rebound (PIR) bursts. Consequently, network activity shifted to bursting rhythms if $g_{syn}$ was sufficiently strong. Since PIR bursting duration does not significantly change with the availability of CaT current (Park et al., 2021), these bursting rhythms tended to be regular and synchronized. On the other hand, when CaL current was increased, STN cell tended to generate longer PIR bursts in response to inhibitory input from GPe cells (Park et al., 2021). This extended response of STN cell forced GPe cells to produce longer bursting rhythms, resulting in low-frequency bursting rhythms in region of synchronized pathological activity. Additionally, a broad band of intermittently synchronized activity patterns emerged, highlighting the important role of CaL current in beta rhythm generation in the STN-GPe network.

The STN-GPe network demonstrates entrainment in response to external periodic inputs into STN cells and this entrainment starts in the low $I_{gpe}$/low $g_{syn}$ area and expands toward the high $I_{gpe}$/high $g_{syn}$ region. The initiation and expansion of entrainment depends on input frequencies, but the overall behaviors remain qualitatively similar. As the entrainment spreads toward the high $I_{gpe}$/high $g_{syn}$ region, the region of intermittently synchronized activities shifts and shrinks accordingly. When CaT current was increased, the low $I_{gpe}$/low $g_{syn}$ region responded even to small input amplitudes; however, the expansion of entrainment region required stronger inputs because the synchronized bursting rhythms resisted the entrainment. Conversely, when CaL current was increased, the region of intermittently synchronized activities tended to persist over a wide range of input amplitudes.

It was suggested (Pan et al., 2016) that STN burst discharges and beta oscillations are driven by two independent mechanisms. They suggested that the indirect pathway (cortex-striatum-GPe-STN-internal pallidum) is responsible for lower beta-band and less bursty beta rhythms, while the hyperdirect pathway (cortex-STN-internal pallidum) accounts for high beta-band and bursty STN burst discharges (Blumenfeld et al., 2017). They demonstrated that a CaT blocker suppressed STN burst discharges and alleviated bradykinesia, whereas the application of hyperpolarizing current (increasing CaT availability) recapitulated bradykinesia and suppressed beta power in LFPs in the STN. This finding aligns with our simulation results. We found (Park et al., 2021) that STN cells generate burst discharges when hyperpolarized and this process was initiated by CaT current and reinforced by CaL current. In the STN-GPe network, increased availability of CaT current resulted in the expansion of synchronized bursting rhythms. Consequently, the region of intermittently synchronized activity was pushed toward the high $I_{gpe}$/high $g_{syn}$ and shrank. The fact that the synchronized bursting rhythms in STN-GPe network resist external inputs suggests that the STN-GPe network does not respond precisely to cortical input. This, in turn, may result in decreased locomotor behavior, potentially due to interrupted information transfer via BG. Considering that bradykinesia is not an early PD symptom, the STN in advanced PD becomes more sensitive to the inhibitory input from GPe, hence the availability of CaT is substantially increased, and this results in STN burst discharges.

On the other hand, (Pan et al., 2016) suggest that beta oscillations have been linked to the dynamics of STN-GPe network in response to cortico-subthalamic transmission. While there aren't many experimental findings on the role of CaL current in PD, our simulation results suggest that CaL currents enrich and sustain beta oscillations in STN-GPe network even under external inputs of moderate strength. Specifically, CaL current enables STN cells to generate longer burst discharges in response to inhibitory input from GPe cells, reinforcing the existence of intermittently synchronized activities. Additionally, the fact that the region of intermittently synchronized activity tends to persist over a wide range of input amplitudes implies that the STN-GPe network does not respond well to the input from cortical areas. This result implicates the significant role of the STN-GPe network in the Parkinsonian BG, as the STN-GPe network can generate beta rhythms autonomously, and these rhythms can be reinforced within the Parkinsonian BG via CaL currents. It is not yet clear how dopamine depletion increases the availability of CaL currents, but our results suggest it to be an important question for future experimental research.

Mechanistic understanding of the properties of the synchronized oscillatory activity in the subthalamo-pallidal networks may assist with improvements of the deep brain stimulation (DBS) of STN in PD. On the one hand, experimental low-frequency (~20Hz) STN DBS is detrimental to motor behaviors since it enhances beta-band synchrony in the BG (Eusebio et al., 2008; Chen et al., 2011; Timmermann and Florin, 2012), probably through some resonant-like properties (Eusebio et al., 2009; Baaske et al., 2020; Zapata Amaya et al., 2023). This effect is similar to our modeling results for the stimulation with a strong magnitude periodic signal. On the other hand, clinical effectiveness of high-frequency STN DBS has been linked to the disruption of pathological beta-band rhythms (see, e.g., Eusebio et al., 2011; Petersson et al., 2020). For more effective DBS procedures, it may be crucial to understand how the cellular properties of neurons contribute to the generation of these pathological network rhythms.

The current study leaves out several important factors that may contribute to the mechanisms underlying synchronized oscillatory activity in the STN-GPe networks of the brain. For example, it would be valuable to know how a more detailed and realistic representation of the intricate cortico-basal ganglia-thalamic circuitry might affect phenomena considered in this study. In the present study, however, we focused exclusively on the STN-GPe network. Within this narrower context, two questions naturally arise: how network size and network organization influence the dynamics. Although we did not investigate these two questions systematically, we conducted limited numerical analyses to explore them. When we increased the network size by 50% while keeping the connectivity scheme unchanged, we observed that the overall results remained qualitatively similar (data not shown). This suggests that the general properties of the intermittent synchronous dynamics are likely to be preserved in larger networks.

Network connectivity, which generally plays an important role in shaping neuronal dynamics, presents a more complex case. Previous investigations of different network architectures suggest that network connectivity can affect the synchronization properties of STN-GPe networks (McLoughlin and Lowery, 2024; see also Terman et al., 2002 who considered different architectures). However, it has long been known that neurons in the basal ganglia circuits exhibit some degree of specificity of the responses to (or correlations with) activation of particular joints. This observation indicates that the topology of the basal ganglia networks is unlikely to be random or all-to-all; rather, it likely reflects a structured spatial organization with substantial local connectivity. Accordingly, this study focused on spatially structured networks with predominantly local connections as it appears to be more anatomically relevant (even though not exclusively possible) configuration for the brain circuit under consideration. We performed limited testing of partially randomized network connections, where each GPe cell inhibits a reciprocally connected STN cell and two randomly selected STN cells. Simulation results showed that the existence of intermittently synchronized regimes persists while the shape of the region of these regimes in the parameter space may change (data not shown). The situation was similar when the network size was increased by a factor of one and a half. Nevertheless, the impact of network connectivity remains an important subject of a future study. Other aspects of neural organization, such as heterogeneity, noise, and synaptic plasticity, have not been incorporated in the present model, and exploring these features may also be pursued in future studies.

Finally, we would like to take a broader view of our results by considering them within a wider context of the factors that influence the network dynamics during the transition from uncorrelated desynchronized activity to more synchronized and correlated patterns. One such factor is the aforementioned network architecture, which has been considered in the context of the STN-GPe networks (McLoughlin and Lowery, 2024) but lies beyond the scope of the present study. The properties of synapses and oscillators also play important roles. In the context of the dopamine dependent parameters that affect cellular and synaptic properties, the synaptic and cellular effects were observed even in very minimal models (e.g., Park and Rubchinsky, 2011, 2012). The present study contributes to this line of works by demonstrating how the interplay among slow calcium currents, external inputs, and excitatory-inhibitory bursting network organization may affect the transition from uncorrelated (presumably healthy) activity to realistic intermittently synchronous patterns, and ultimately to strongly correlated dynamics.

**Acknowledgement**
This work was partially supported by AMS-Simons Grant for PUI (SA, CP) and NSF DMS 2401922 (CP).

**Author Declarations**
**Conflict of Interest**
The authors declare that they have no known competing financial interests or personal relationships that could have appeared to influence the work reported in this paper.

**Author Contributions**
C.P. and S.A conceived the study, set up the model, and conducted the numerical simulation. C.P, S.A, and L.R analyzed the data and wrote the manuscript. All authors reviewed and approved the final manuscript.

**Data Availability**
Data sharing is not applicable to this article as no new data were created or analyzed in this study.